%% file: main.tex
\documentclass[conference, a4paper]{IEEEtran}
\IEEEoverridecommandlockouts

\usepackage{cite}
\usepackage{algorithm}
\usepackage[noend]{algpseudocode}
\usepackage{amsmath,amssymb,amsfonts}
\usepackage[pdftex]{graphicx}
\usepackage{subcaption}
\usepackage{textcomp}
\usepackage{xcolor}
\usepackage{pifont}
\usepackage{multirow}
\usepackage{makecell}
\usepackage{array}
\usepackage{diagbox}
\usepackage[left=1.42cm,right=1.24cm,top=1.8cm,bottom=4.30cm]{geometry}

\begin{document}

\title{Robust Semantic Communications for Speech Transmission}
\author{Zhenzi Weng\IEEEauthorrefmark{1}, Zhijin Qin\IEEEauthorrefmark{2}, and Geoffrey Ye Li\IEEEauthorrefmark{1}\\
\small \IEEEauthorrefmark{1} Department of Electrical and Electronic Engineering, Imperial College London, London, UK\\
\small \IEEEauthorrefmark{2} Department of Electronic Engineering, Tsinghua University, Beijing, China\\
\small Email:  z.weng@imperial.ac.uk, qinzhijin@tsinghua.edu.cn, geoffrey.li@imperial.ac.uk
\thanks{The work was supported in part by the National Key Research and Development Program of China under Grant 2023YFB2904300; in part by the National Natural Science Foundation of China (NSFC) under Grant 62293484; and in part by the Beijing National Research Center for Information Science and Technology (BNRist), Beijing, China.}
}

\maketitle
\input{Section/abstract}

\IEEEpeerreviewmaketitle

\input{Section/introduction}

\input{Section/system_model.tex}

\input{Section/proposed_system}

\input{Section/numerical_results}

\input{Section/conclusion}

\bibliographystyle{IEEEtran}
\bibliography{reference.bib}

\end{document}

%% file: Section/abstract.tex
\begin{abstract}
In this paper, we propose a robust semantic communication system for speech transmission, named Ross-S2T, by delivering the essential semantic information. Specifically, we consider the speech-to-text translation (S2TT) as the transmission goal. First, a new deep semantic encoder is developed to convert speech in the source language to textual features associated with the target language, facilitating the end-to-end semantic exchange to perform the S2TT task and reducing the transmission data without performance degradation. To mitigate semantic impairments inherent in the corrupted speech, a novel generative adversarial network (GAN)-enabled deep semantic compensator is established to estimate the lost semantic information within the speech and extract deep semantic features simultaneously, which enables robust semantic transmission for corrupted speech. Furthermore, a semantic probe-aided compensator is devised to enhance the semantic fidelity of recovered semantic features and improve the understandability of the target text. According to simulation results, the proposed Ross-S2T exhibits superior S2TT performance compared to conventional approaches and high robustness against semantic impairments.

\end{abstract}


%% file: Section/introduction.tex
\section{Introduction}
Semantic communications have been regarded as a promising solution to tackle the technical challenges in conventional communication systems and have attracted significant research attention in recent years~\cite{9770094}. The advancement of semantic communications derives from the ability to explore semantic information and achieve semantic exchange, which revolutionizes many aspects of wireless communications~\cite{10639525}.

According to the three-level communication architecture proposed by Shannon and Weaver~\cite{weaver1953recent}, semantic communications are the second level of communications that prioritize conveying the underlying meaning by representing the input message with minimal ambiguity through semantics, which overcomes the limitation of conventional communications to process data at the bit level and drives the evolution of intelligent communications. However, there is no consensus on the definition of semantics, hindering the representation of semantic information with a rigorous formula. Thanks to the thriving of artificial intelligence (AI) in diverse areas, deep learning (DL)-enabled semantic communication paradigm breaks the bottleneck of a mathematical theory to quantify the semantics and has shown its great potential to learn semantic information through sophisticated neural networks (NNs). DL-enabled semantic communications have experienced unprecedented developments due to the ubiquity of intelligent mobile devices and the booming demand for semantic-driven data transmission.

Particularly, the pioneering work on DL-enabled semantic communications, named DeepSC~\cite{9398576}, has been proposed to recover the accurate text. A variant of DeepSC, named R-DeepSC~\cite{10486856}, has been devised to eliminate semantic noise and facilitate robust text transmission. In~\cite{9450827}, Weng~\emph{et al.} introduced a semantic communication system for speech transmission, named DeepSC-S, to extract and transmit global semantic features. Inspired by DeepSC-S, a deep speech semantic transmission scheme has been developed in~\cite{10094680} by adopting a flexible rate-distortion trade-off to achieve end-to-end (E2E) optimization. Huang~\emph{et al.}~\cite{9953076} considered a semantic communication system for image transmission. Jiang~\emph{et al.}~\cite{9955991} presented a video semantic transmission paradigm.

Furthermore, Xie~\emph{et al.}~\cite{9830752} established a task-oriented semantic communication system for machine translation and visual question-answering tasks by fusing textual and visual semantic features. Weng~\emph{et al.}~\cite{10038754} designed a speech semantic transmission scheme, named DeepSC-ST, to perform speech recognition and synthesis tasks, and further explored the speech-to-text translation (S2TT) and speech-to-speech translation (S2ST) tasks in~\cite{10333632} by incorporating a machine translation module into DeepSC-ST. In~\cite{9796572}, Xu~\emph{et al.} proposed reinforcement learning-enabled semantic communications for scene classification in unmanned aerial systems. Zhang~\emph{et al.} investigated a semantic communication system for extended reality (XR) tasks by transmitting highly compressed semantic information to reduce network traffic. In~\cite{10431795}, a unified multimodal multi-task semantic communication architecture, named U-DeepSC, has been developed by sharing trainable parameters amongst various tasks to reduce the redundancy of semantic features and accelerate the inference process.

In this paper, a robust semantic communication system for speech transmission, named Ross-S2T, is proposed. We argue that existing works on task-oriented semantic communications for speech only extract textual semantic features constrained to the source language, i.e., shallow semantic features, encouraging the exploration of deep semantic features spanning various languages. Moreover, we investigate the intractable semantic impairments. In this context, the S2TT task is considered in semantic transmission scenarios with corrupted speech input. The contributions of this paper are summarized as follows:
\begin{itemize}
\item A semantic communication system for S2TT, named DeepSC-S2T, is developed to obtain the deep semantic features associated with the target language by leveraging a semantic extractor and a novel semantic converter.

\item According to our comprehensive literature review, speech semantic impairments are not investigated in semantic communications. We propose a generative adversarial network (GAN)-enabled deep semantic compensator to estimate the damaged information in the corrupted speech and generate deep semantic features simultaneously.

\item To further reduce the semantic loss at the recovered features, a semantic impairment probe-aided compensator is established to perceive and calibrate the corrupted semantic features at the receiver, thereby improving the accuracy of the produced target text.

\item Simulation results verify the superiority of the DeepSC-S2T to serve the S2TT task and the robustness of the Ross-S2T to contend with semantic impairments.

\end{itemize}



%% file: Section/system_model.tex
\section{System Model}
This section introduces robust semantic communication systems for speech transmission and considers S2TT the transmission goal. The system aims to address two primary challenges. The first challenge is to deliver E2E semantic exchange and achieve efficient transmission from speech in the source language to text in the target language. The second is to devise a semantic impairment suppression mechanism to reduce semantic impairments within the corrupted speech. To this end, the novel deep semantic coding mechanism is established to facilitate speech transmission for S2TT, and the deep semantic compensator is first developed to compensate for the sophisticated semantic impairments.

\subsection{Clear Speech Input}
\begin{figure}[tbp]
\includegraphics[width=0.487\textwidth]{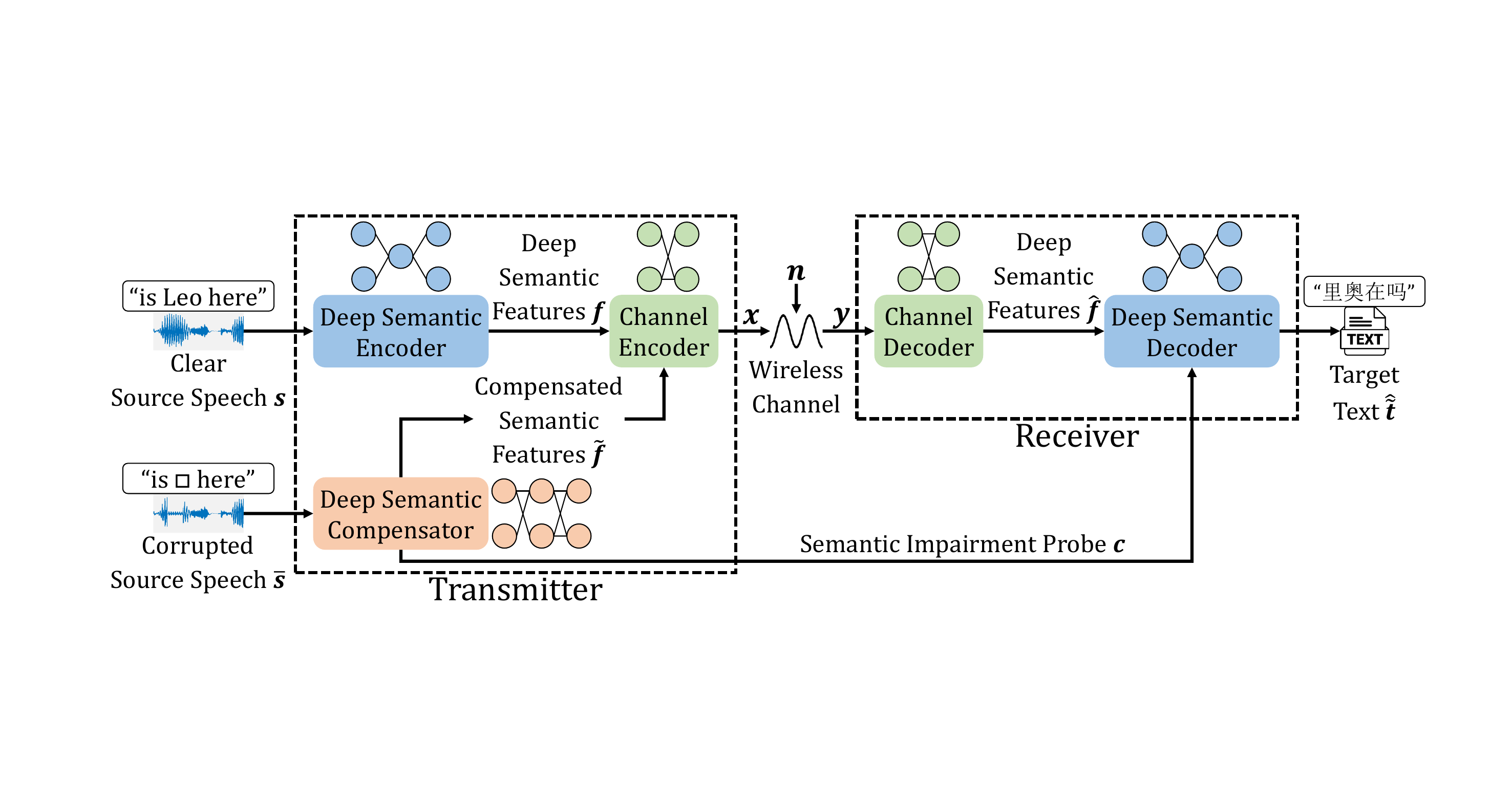}
\centering 
\caption{Model structure of robust semantic communications for speech-to-text translation.}
\label{system model}
\end{figure}

\begin{figure}[tbp]
\includegraphics[width=0.3\textwidth]{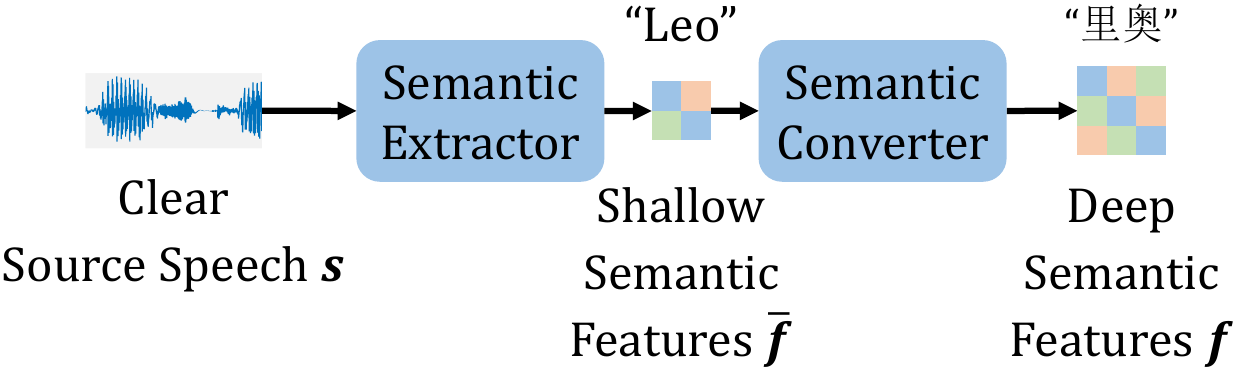}
\centering 
\caption{The proposed deep semantic encoder.}
\label{deep semantic encoder module}
\end{figure}

The designed system is tailored for communication scenarios involving transmitter and receiver users from different linguistic backgrounds and facilitates the conversion of multimodal data from speech to text. The framework of robust semantic communications for S2TT is shown in Fig.~\ref{system model}. From the figure, when the system input is clear speech, the deep semantic encoder compresses the speech sequence, $\boldsymbol s$, and produces the deep semantic features, $\boldsymbol f$. Note that the existing work on semantic communications for S2TT~\cite{10333632} only extracts the shallow semantic features related to the source language. It also generates intermediate source text at the receiver before producing the final target text, which hinders E2E training and requires additional computational resources for the machine translation module. To resolve this issue, we devise a novel semantic converter to transform shallow semantic features into deep semantic features, as shown in Fig.~\ref{deep semantic encoder module}. From the figure, $\boldsymbol f$ can be expressed according to $\boldsymbol s$ as follows,
\begin{equation}
\boldsymbol f={\mathfrak T}_{\mathrm{SC}}({\mathfrak T}_{\mathrm{SE}}(\boldsymbol s))\;\;\;\mathrm w.\mathrm r.\mathrm t.\;\;\;\boldsymbol\alpha,
\label{deep semantic encoder}
\end{equation}
where ${\mathfrak T}_{\mathrm{SE}}(\cdot)$ and ${\mathfrak T}_{\mathrm{SC}}(\cdot)$ are the semantic extractor and the semantic converter, respectively. ${\mathfrak T}_{\mathrm{DS}}=({\mathfrak T}_{\mathrm{SE}},\;{\mathfrak T}_{\mathrm{SC}})$. $\boldsymbol\alpha$ is all trainable NN parameters of ${\mathfrak T}_{\mathrm{DS}}$.

The channel encoder maps $\boldsymbol f$ to the symbols, $\boldsymbol x$, before transmission over the wireless channel, denoted as,
\begin{equation}
\boldsymbol x={\mathfrak T}_{\mathrm C}(\boldsymbol f)\;\;\;\mathrm w.\mathrm r.\mathrm t.\;\;\;\boldsymbol\beta,
\label{transmitter}
\end{equation}
where ${\mathfrak T}_{\mathrm C}(\cdot)$ is the channel encoder and $\boldsymbol\beta$ is its NN parameters.

The encoded symbols, $\boldsymbol x$, are affected by the channel fading and channel noise after passing through the wireless channel layer, and the received symbols, $\boldsymbol y$, can be written as
\begin{equation}
\boldsymbol y=\boldsymbol h\ast\boldsymbol x+\boldsymbol n,
\label{channel}
\end{equation}
where $\boldsymbol h$ represents the fading channel and $\boldsymbol n$ is the additive white Gaussian noise (AWGN).

At the receiver, the target text, $\widehat{\widetilde{\boldsymbol t}}$, is obtained by feeding the recovered features, $\widehat{\boldsymbol f}$, into the deep semantic decoder, which can be modelled as
\begin{equation}
\widehat{\widetilde{\boldsymbol t}}=\mathfrak T_{\mathrm{DS}}^{-1}(\mathfrak T_{\mathrm C}^{-1}\left(\boldsymbol y\right))\;\;\;\;\mathrm w.\mathrm r.\mathrm t.\;\;\;\boldsymbol\rho,
\label{receiver}
\end{equation}
where $\mathfrak T_{\mathrm C}^{-1}(\cdot)$ and $\mathfrak T_{\mathrm{DS}}^{-1}(\cdot)$ denotes the channel decoder and the deep semantic decoder, respectively. $\boldsymbol\rho$ represents their trainable NN parameters.

\subsection{Corrupted Speech Input}
In practical communication systems, the integrity of input speech is highly susceptible to perturbations induced by surrounding environments or unstable network connections, resulting in the potential degradation of original speech. In this work, our endeavors towards semantic communications for S2TT extend to the corrupted speech input, wherein some semantic information within the speech is inaccessible due to the introduction of semantic impairments. Semantic impairments refer to external interference that damages the integrity of speech information. As shown in Fig.~\ref{system model}, the corrupted speech, $\overline{\boldsymbol s}$, contains limited semantic information. Particularly, the deep semantic compensator tasks $\overline{\boldsymbol s}$ as input and generates the compensated deep semantic features, $\widetilde{\boldsymbol f}$, which predicts the lost information and extracts textual semantic features in the target language simultaneously, written as
\begin{equation}
\widetilde{\boldsymbol f}={\mathfrak T}_{\mathrm{DC}}(\overline{\boldsymbol s})\;\;\;\mathrm w.\mathrm r.\mathrm t.\;\;\;\boldsymbol\delta,
\label{deep semantic compensator}
\end{equation}
where ${\mathfrak T}_{\mathrm{DC}}(\cdot)$ indicates the deep semantic compensator and $\boldsymbol\delta$ is the corresponding trainable NN parameters.

Furthermore, a semantic impairment probe, $\boldsymbol c$, containing an index vector with position information of the corrupted semantic features, is attained and transmitted to the receiver over a reliable channel. The motivation behind the semantic impairment probe is to strengthen the semantic fidelity of the recovered deep semantic features by further reducing the semantic ambiguity caused by corrupted semantic features.

%% file: Section/proposed_system.tex
\section{Robust Semantic Communications for Speech Transmission}
To address the preceding challenges, we adopt a two-stage training scheme. Particularly, a semantic transmission paradigm for S2TT based on clear speech input, named DeepSC-S2T, is first proposed. Then, a dual-compensator mechanism is developed to enhance robust semantic communications, named Ross-S2T, which utilizes a GAN-enabled deep semantic compensator and a semantic impairment probe-aided compensator to acquire as accurate deep semantic features as possible at the receiver.

\subsection{DeepSC-S2T}
\begin{figure}[tbp]
\begin{minipage}[t]{1.0\linewidth}
\centering
\includegraphics[width=1.0\textwidth]{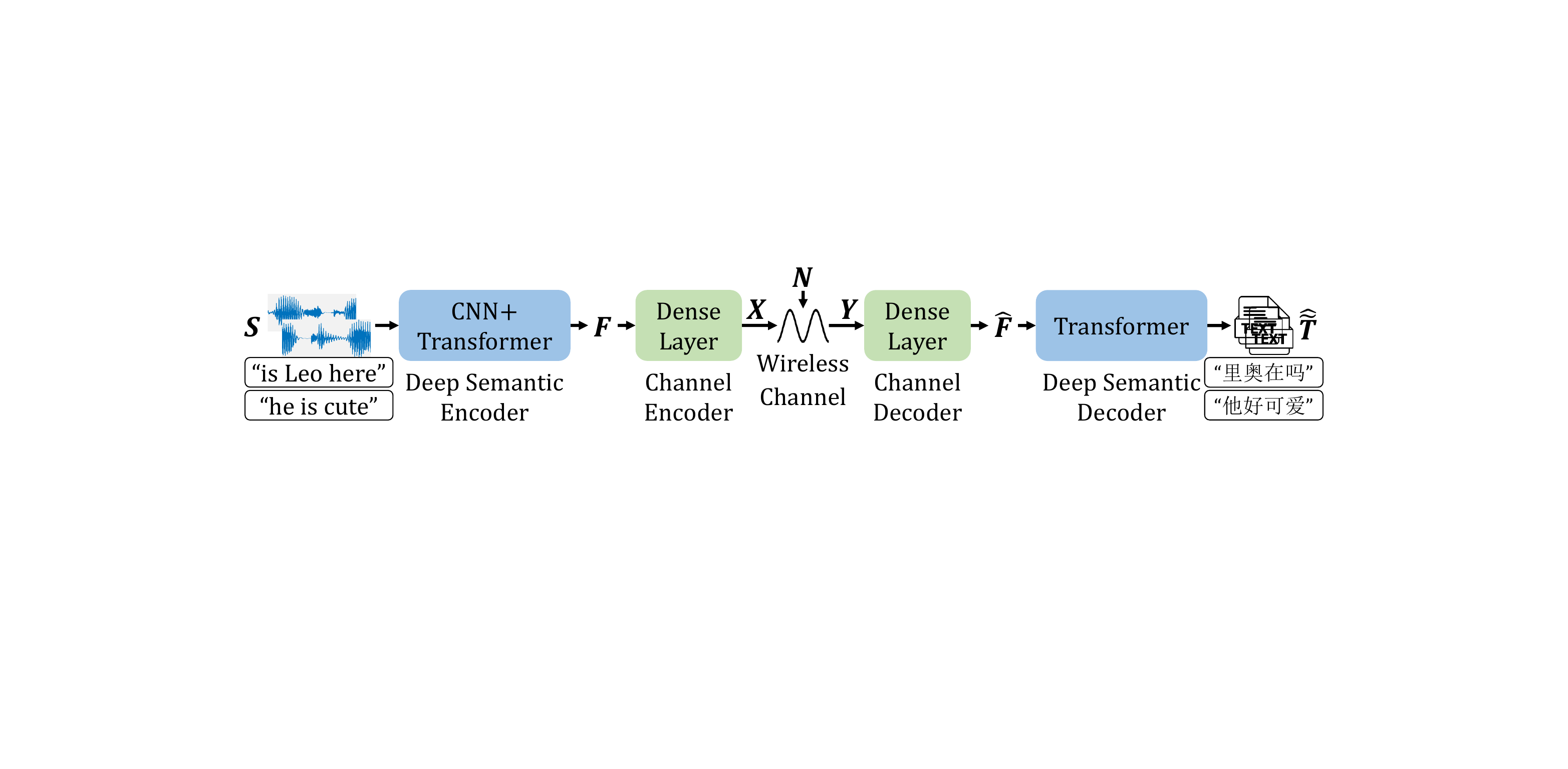}
\subcaption{First training stage: Train DeepSC-S2T with clear speech input.}
\label{first training stage}
\vspace{0.1cm}  
\end{minipage}
\begin{minipage}[t]{1.0\linewidth}
\centering
\includegraphics[width=1.0\textwidth]{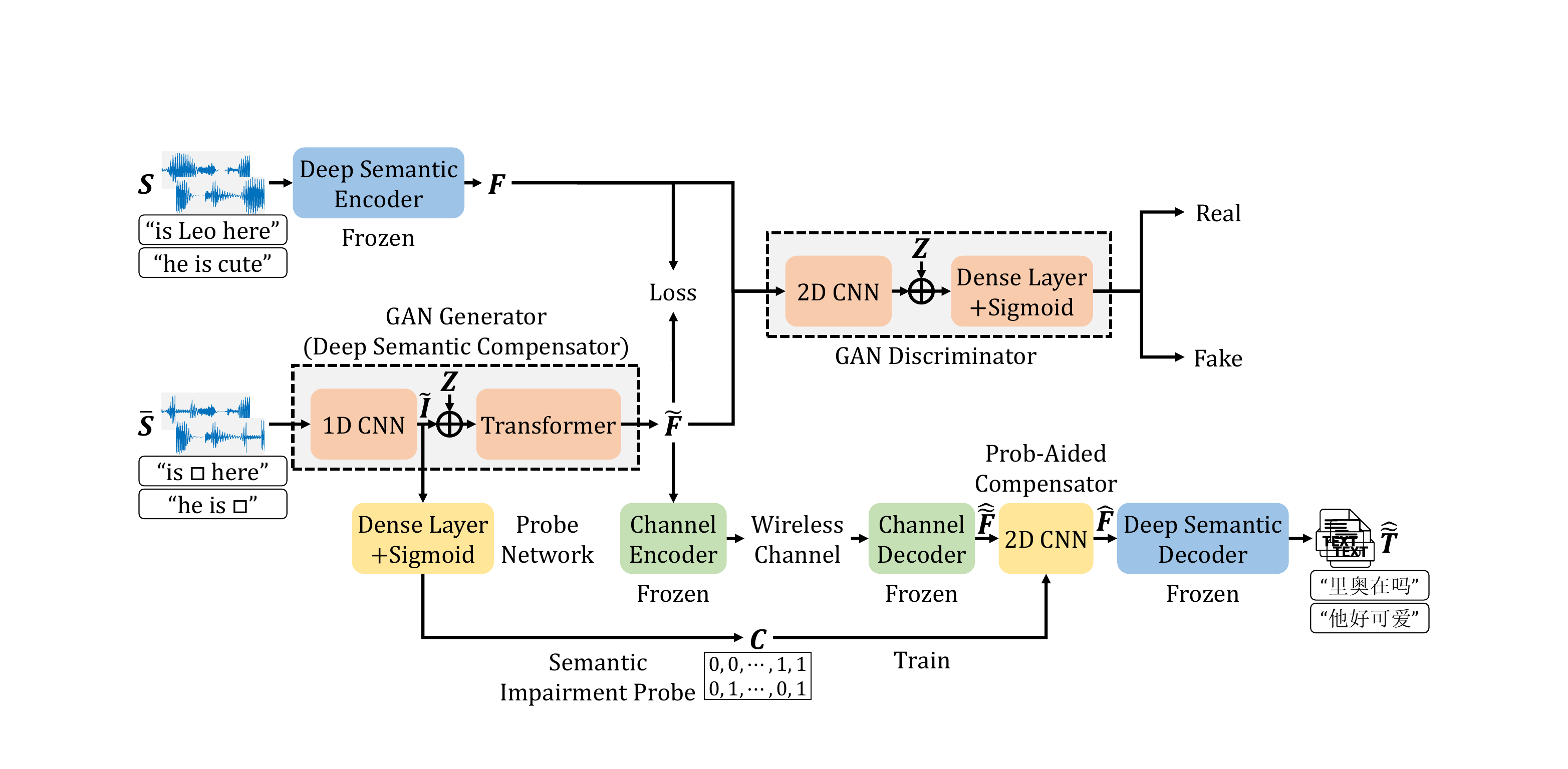}
\subcaption{Second training stage: Train Ross-S2T with corrupted speech input.}
\label{second training stage}
\vspace{0.1cm}  
\end{minipage}
\begin{minipage}[t]{1.0\linewidth}
\centering
\includegraphics[width=1.0\textwidth]{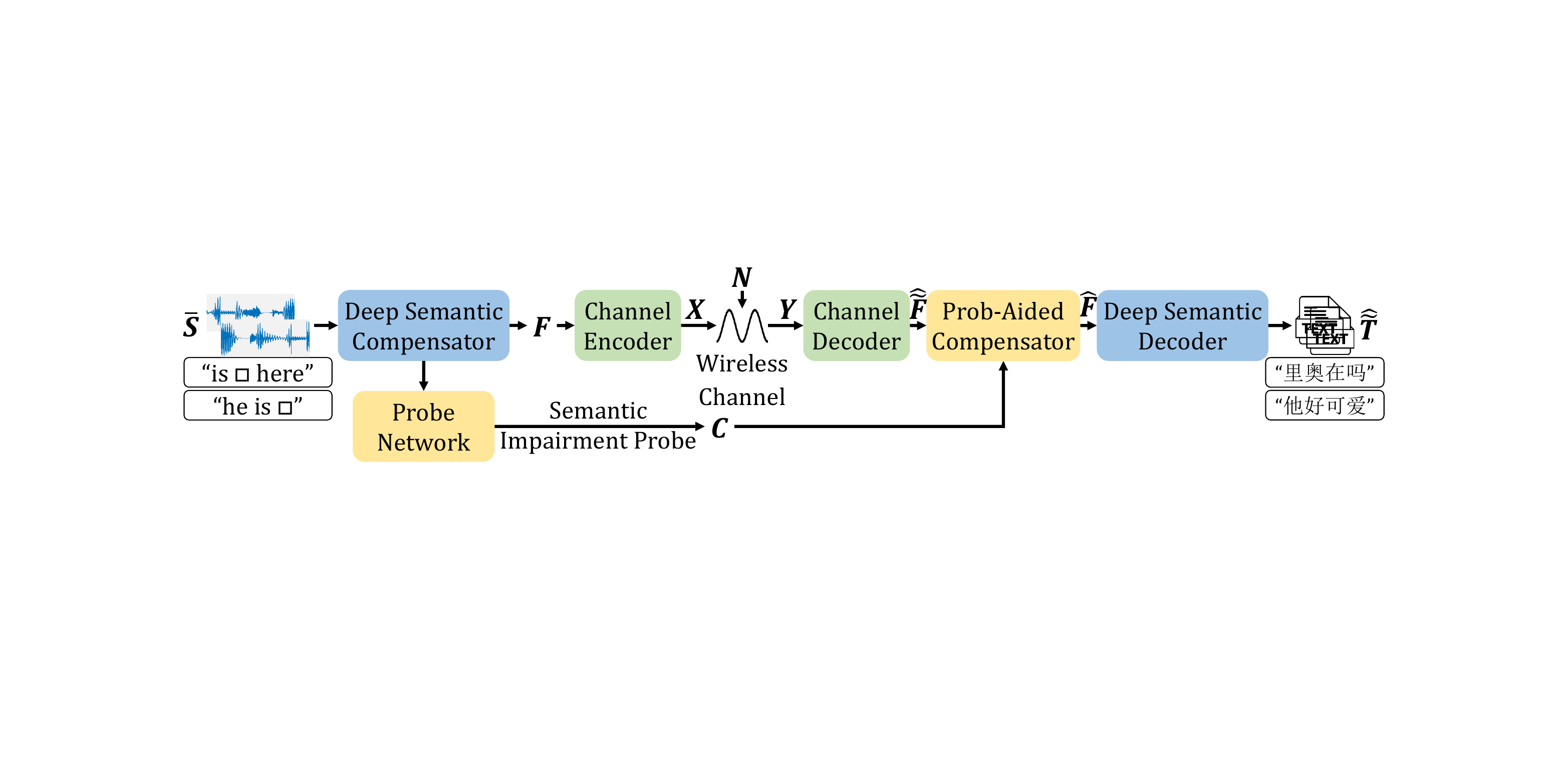}
\subcaption{Testing stage: Test Ross-S2T with corrupted speech input.}
\label{testing stage}
\end{minipage}
\caption{Model structure of Ross-S2T for robust semantic communications with clear and corrupted speech inputs.}
\label{proposed system}
\end{figure}

The proposed Ross-S2T is shown in Fig.~\ref{proposed system}. From the figure, at the first training stage, the convolutional neural network (CNN) module condenses the clear speech and the transformer module further extracts the features, $\boldsymbol F$, before passing through the dense layer-enabled channel encoder to attain symbols, $\boldsymbol X$. The dense layer constructs the channel decoder to process the receiver symbols, $\boldsymbol Y$, and the transformer-enabled deep semantic decoder is leveraged to produce multiple target text sequences, $\widehat{\widetilde{\boldsymbol T}}$. To boost the efficient semantic transmission for serving the S2TT task, the label-smoothing regularization-aided cross-entropy (LSR-CE) is adopted as the E2E loss function to train the DeepSC-S2T, which is expressed as
\begin{equation}
{\mathcal L}_{\mathrm{LSR-CE}}(\widetilde{\boldsymbol T},\;\widehat{\widetilde{\boldsymbol T}};\;\boldsymbol\theta)=\kappa{\mathcal L}_{\mathrm{CE}}+\sum_{\widetilde l=1}^{\widetilde L}f_w\left(w_e\right),
\label{LSR-CE loss}
\end{equation}
where $\kappa\in\left[0,1\right]$ is a hyperparameter that signifies the confidence level associated with the predicted tokens in $\widehat{\widetilde{\boldsymbol T}}$ matching the true tokens in $\widetilde{\boldsymbol T}$. Note that $\widetilde{\boldsymbol T}$ is the accurate text sequence in the target language, $\widetilde{\boldsymbol T}$. The trainable parameters $\boldsymbol\theta=(\boldsymbol\alpha,\;\boldsymbol\beta,\;\boldsymbol\rho)$. $\widetilde L$ represents the number of tokens in $\widetilde{\boldsymbol T}$. ${\mathcal L}_{\mathrm{CE}}$ is the CE loss. The token $w_e$ belongs to a vocabulary group containing $E$ tokens and $w_e\neq{\widetilde t}_{\widetilde l}$. Additionally, $f_w\left(w_e\right)$ describes the confidence level associated with ${\widehat{\widetilde t}}_{\widetilde l}$ matching $w_e$ instead of ${\widetilde t}_{\widetilde l}$, which can be written as
\begin{equation}
f_w\left(w_e\right)=-\sum_{e=1,w_e\neq{\widetilde t}_{\widetilde l}}^E\frac\kappa{E-1}p(w_e)\log p({\widehat{\widetilde t}}_{\widetilde l}).
\label{CE loss 2}
\end{equation}

The intuition behind loss ${\mathcal L}_{\mathrm{LSR-CE}}$ is to introduce a level of confusion in predicting the target text, which enables training uncertainty but ultimately improves prediction accuracy in the testing stage.

\subsection{Ross-S2T}
As aforementioned, the deep semantic compensator is responsible for estimating the damaged semantic information in $\overline{\boldsymbol S}$, extracting the deep semantic features with the least dissimilarity to $\boldsymbol F$, and returning the semantic impairment probe matrix to record the positional information of corrupted deep semantic features. In the second training stage, we propose a dual-compensator mechanism, including the GAN-enabled deep semantic compensator at the transmitter and the probe-aided compensator at the receiver. Particularly, the trained deep semantic encoder is leveraged to obtain $\boldsymbol F$ as the real data for the discriminator. The generator is developed to process the corrupted speech, $\overline{\boldsymbol S}$, and generate the fake data $\widetilde{\boldsymbol F}$ to fool the discriminator by adopting the 1D CNN module followed by the latent space, $\boldsymbol Z$, and the transformer module. Note that the intermediate semantic representation, $\widetilde{\boldsymbol I}$, are attained as the output of the 1D CNN module. The discriminator distinguishes whether the input data is real or fake, incorporating the 2D CNN module before latent space $\boldsymbol Z$ followed by the dense and sigmoid layers. Denote the trainable NN parameters of the discriminator and the generator as ${\boldsymbol\gamma}_{\mathrm D}$ and ${\boldsymbol\gamma}_{\mathrm G}$, respectively, i.e., $\boldsymbol\gamma=({\boldsymbol\gamma}_{\mathrm D},\;{\boldsymbol\gamma}_{\mathrm G})$. Then, the loss function adopted for training the discrimination can be expressed as
\begin{equation}
{\mathcal L}_{\mathrm D}(\boldsymbol S,\;\overline{\boldsymbol S})=\frac12\left({\mathfrak T}_{\mathrm D}\left({\mathfrak T}_{\mathrm{DS}}\left(\boldsymbol S\right)\right)-1\right)^2+\frac12\left({\mathfrak T}_{\mathrm D}({\mathfrak T}_{\mathrm G}(\overline{\boldsymbol S}))\right)^2,
\label{discrimination loss}
\end{equation}
where ${\mathfrak T}_{\mathrm{D}}(\cdot)$ is the discriminator and ${\mathfrak T}_{\mathrm{G}}(\cdot)$ is the generator.

The ${\boldsymbol\gamma}_{\mathrm G}$ can be updated as follows,
\begin{equation}
\begin{split}
{\mathcal L}_{\mathrm G}(\boldsymbol S,\;\overline{\boldsymbol S})&=\frac12\xi{\mathcal L}_{\mathrm{MSE}}+\frac12{\mathcal L}_{\mathrm{ADV}} \\
&=\frac12\xi\left(\boldsymbol F-\widetilde{\boldsymbol F}\right)^2+\frac12\left({\mathfrak T}_{\mathrm D}({\mathfrak T}_{\mathrm G}(\overline{\boldsymbol S}))-1\right)^2,
\end{split}
\label{generator loss}
\end{equation}
where $\xi$ is a hyperparameter to balance the weights of the MSE loss, ${\mathcal L}_{\mathrm{MSE}}$, and the adversarial loss, ${\mathcal L}_{\mathrm{ADV}}$.

The trained generator aims to create realistic data to deceive the discriminator and produce compensated deep semantic features, $\widetilde{\boldsymbol F}$, that closely resemble $\boldsymbol F$.

Furthermore, a probe network, $\mathfrak T_{\mathrm{PN}}(\cdot)$, is established at the transmission,  taking $\widetilde{\boldsymbol I}$ as the input and providing the semantic impairment probe, $\boldsymbol C$. The loss function for training the $\mathfrak T_{\mathrm{PN}}(\cdot)$ is denoted as
\begin{equation}
{\mathcal L}_{\mathrm{PN}}(\boldsymbol I,\;\widetilde{\boldsymbol I},\;\boldsymbol C;\;\boldsymbol\varepsilon)=\sum_{l^{'}=1}^{L^{'}}\left(i_{l^{'}}-c_{l^{'}}{\widetilde i}_{l^{'}}\right)^2,
\label{Ross-S2T: probe network loss}
\end{equation}
where $\boldsymbol I$ is the intermediate semantic representation extracted from the clear speech. $\boldsymbol\varepsilon$ is the NN parameters of $\mathfrak T_{\mathrm{PN}}(\cdot)$.

To further enhance the fidelity of the received deep semantic features, $\widehat{\widetilde{\boldsymbol F}}$, the learned semantic impairment probe, $\boldsymbol C$, is utilized to identify the corrupted deep semantic features in $\widehat{\widetilde{\boldsymbol F}}$ and the CNN-enabled probe-aided compensator commits to reducing semantic errors between the identified corrupted features and the corresponding accurate features. Denote the NN parameters of the probe-aided compensator is $\boldsymbol\zeta$ and it can be updated by
\begin{equation}
{\mathcal L}_{\mathrm{PC}}(\widetilde{\boldsymbol T},\;\widehat{\widetilde{\boldsymbol T}},\;\boldsymbol C;\;\boldsymbol\zeta)=-\sum_{l^\backprime=1,\;c_{l^\backprime}\neq0}^{L^\backprime}p({\widetilde t}_{l^\backprime})\log p({\widehat{\widetilde t}}_{l^\backprime}),
\label{Ross-S2T: probe-aided compensator loss}
\end{equation}
where $l^\backprime$ indicates the position corresponding to the semantic impairment probe with the value of one.
By introducing the probe-aided compensator, the understandability of the target text can be improved compared to scenarios where the received features are directly fed into the deep semantic decoder. 

As shown in Fig.~\ref{proposed system} (c), the trained GAN generator and probe network are invoked to acquire features $\widetilde{\boldsymbol F}$ and semantic impairment probe $\boldsymbol C$, respectively, under circumstances of corrupted speech input. The probe-aided compensator calibrates the corrupted deep semantic features, and the deep semantic decoder generates the text in the target language.

%% file: Section/numerical_results.tex
\section{Numerical Results}
In the experiments, the corpus~\emph{CoVoST 2} is used as the clear speech dataset. To create the corrupted speech dataset, the clear speech is engulfed by semantic impairments, including background noise and external speech interference. 
The semantic textual similarity (STS)~\cite{agirre-etal-2014-semeval} is adopted to evaluate the performance of the proposed Ross-S2T over wireless channels with the accurate channel state information (CSI). Moreover, we choose English as the source language and Chinese as the target language.

\subsection{Simulation Settings}
\begin{figure}[tbp]
\centering
\begin{minipage}[t]{0.493\linewidth}
\centering
\includegraphics[width=1.0\textwidth]{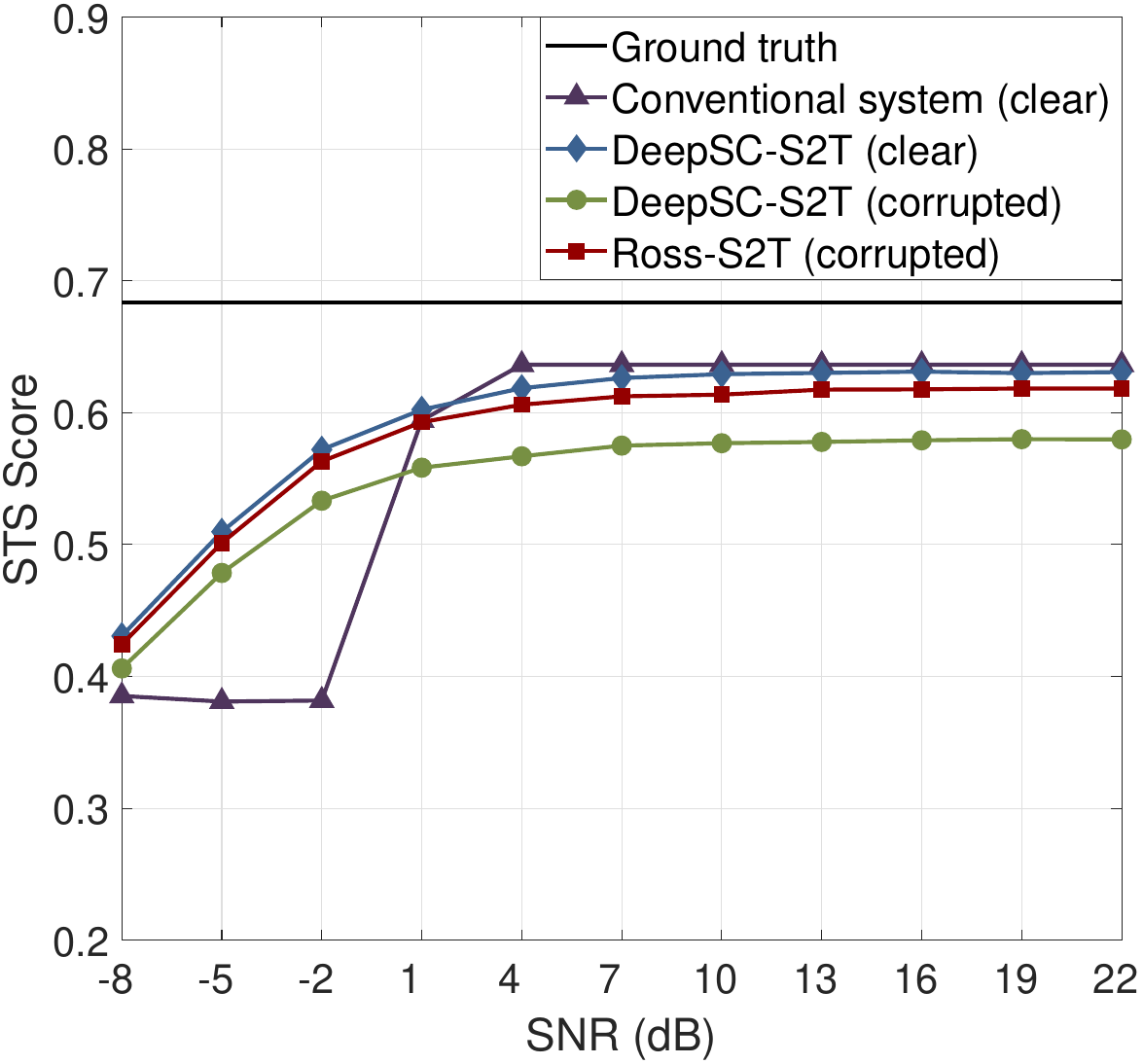}
\subcaption{AWGN channels}
\label{STS AWGN channels}
\end{minipage}
\begin{minipage}[t]{0.493\linewidth}
\centering
\includegraphics[width=1.0\textwidth]{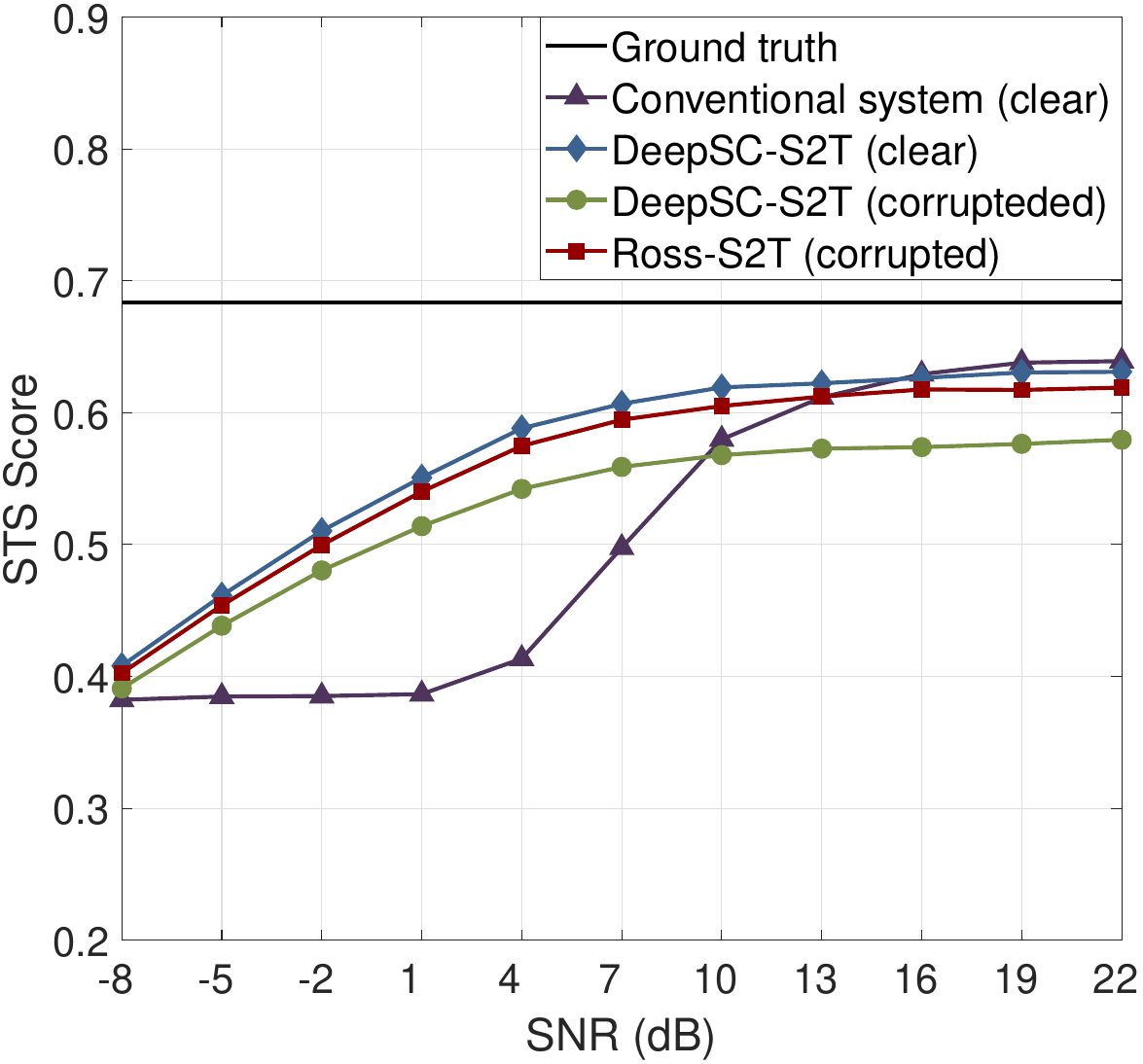}
\subcaption{Rayleigh channels}
\label{STS Rayleigh channels}
\end{minipage} 
\caption{Simulation results of STS scores.}
\label{STS result}
\end{figure}

In the DeepSC-S2T, the semantic extractor includes seven CNN modules, and the semantic converter consists of 12 transformer modules and three CNN modules. The channel encoder/decoder has two dense layers with 1024 units. Eight transformer modules are utilized in the deep semantic decoder. Moreover, the generator of the Ross-S2T is constructed by seven CNN modules, two dense layers, 12 transformer modules, and three CNN modules. The discriminator involves five CNN modules and one dense layer. The probe network includes three dense layers and the probe-aided compensator has five CNN modules. The hyperparameters $\kappa=0.95$ and $\xi=10$.
The STS results of the Ross-S2T are shown in Fig.~\ref{STS result}, where the ground truth results are obtained by feeding the clear speech into an S2TT pipeline constructed from the conformer and the BART. A benchmark is provided by the conventional speech transmission system consisting of the adaptive multi-rate code, the polar code, and the 16-QAM. From the figure, the Ross-S2T attains the STS score of over 0.5 under the Rayleigh channels when SNR$=$1 dB, while the STS score of the conventional system falls below 0.4. Moreover, the DeepSC-S2T manifests a significantly inferior capability in recovering the impaired semantic information within the corrupted speech compared to the proposed Ross-S2T, verifying the superiority of the developed dual-compensator mechanism.

%% file: Section/conclusion.tex
\section{Conclusions}
In this paper, we study the robust semantic communications for speech transmission, named Ross-S2T, to support end-to-end speech-to-text translation (S2TT). Particularly, a deep semantic encoder is developed to learn textual semantic features related to another language from the clear speech, which enables the deep semantic exchange to achieve S2TT at the receiver. Moreover, a generative adversarial network (GAN)-enabled deep semantic compensator and a probe-aided compensator are tailored for corrupted speech scenarios by estimating the impaired semantic information and attaining as accurate deep semantic features as possible. Simulation results demonstrated the superiority of the Ross-S2T in serving the S2TT task and suppressing the semantic impairments.